# A PRG for Lipschitz Functions of Polynomials with Applications to Sparsest Cut

Daniel Kane  
Stanford  
dankane@math.stanford.edu

Raghu Meka  
IAS and DIMACS  
raghu@ias.edu

## Abstract

We give improved pseudorandom generators (PRGs) for *Lipschitz functions* of low-degree polynomials over the hypercube. These are functions of the form $\psi(P(x))$, where $P : \{1, -1\}^n \to \mathbb{R}$ is a low-degree polynomial and $\psi : \mathbb{R} \to \mathbb{R}$ is a function with small Lipschitz constant. PRGs for *smooth* functions of low-degree polynomials have received a lot of attention recently and play an important role in constructing PRGs for the natural class of polynomial threshold functions. In spite of the recent progress, no nontrivial PRGs were known for fooling Lipschitz functions of degree $O(\log n)$ polynomials even for constant error rate. In this work, we give the first such generator obtaining a seed-length of $(\log n)\tilde{O}(\ell^2/\varepsilon^2)$ for fooling degree $\ell$ polynomials with error $\varepsilon$. Previous generators had an exponential dependence on the degree $\ell$.

We use our PRG to get better integrality gap instances for sparsest cut, a fundamental problem in graph theory with many applications in graph optimization. We give an instance of uniform sparsest cut for which a powerful semi-definite relaxation (SDP) first introduced by Goemans and Linial and studied in the seminal work of Arora, Rao and Vazirani [ARV] has an integrality gap of $\exp(\Omega((\log \log n)^{1/2}))$. Understanding the performance of the Goemans-Linial SDP for uniform sparsest cut is an important open problem in approximation algorithms and metric embeddings. Our work gives a near-exponential improvement over previous lower bounds which achieved a gap of $\Omega(\log \log n)$ [DKSV, KR]. Our gap instance builds on the recent *short code* gadgets of Barak et al. [BGH+].

## 1 Introduction

We study the natural question of constructing pseudorandom generators (PRGs) for functions of low-degree polynomials over the hypercube. For instance, an important class of such functions is polynomial threshold functions (PTFs). These are functions of the form $f : \{1, -1\}^n \to \{1, -1\}$ given as $f(x) = \text{sign}(P(x))$ where $P$ is a real-valued multivariate polynomial. PTFs are an important class of functions with a variety of applications in complexity theory [Bei], learning theory [KS2], voting theory [ABFR] and more. Recently, a lot of attention has been given to the problem of constructing PRGs for PTFs – [DGJ+, RS1, DKN, MZ, Kan1, Kan2]. At the core of most of these results is a PRG that fools *smooth* functions of polynomials. For a function $\psi : \mathbb{R} \to \mathbb{R}$, let $\|\psi\|_{Lip} = \sup_{x \neq y} |\psi(x) - \psi(y)|/|x - y|$.

**Definition 1.1.** *We say a generator $G : \{0, 1\}^r \to \{1, -1\}^n$ $\varepsilon$-fools Lipschitz functions of degree $\ell$ polynomials if for every function $\psi : \mathbb{R} \to \mathbb{R}$ and degree $\ell$ polynomial $P$ with $\|P\| = 1$[1],*

$$\left| \mathbb{E}_{y \sim \{0,1\}^r}[\psi(P(G(y)))] - \mathbb{E}_{x \sim \{1,-1\}^n}[\psi(P(x))] \right| < \varepsilon \cdot \|\psi\|_{Lip}.$$

Non-explicitly there exist generators as above with seed-length $r = O(\log n + \ell \log(1/\varepsilon))$. In this work, we give the best known explicit PRG for Lipschitz functions of polynomials:

**Theorem 1.2.** *There exists an explicit generator $G : \{0, 1\}^r \to \{1, -1\}^n$ that fools Lipschitz functions of degree $\ell$ polynomials with error $\varepsilon$ and has seed-length $r = O((\log n + \log^2(\ell/\varepsilon)) \cdot \ell^2/\varepsilon^2)$.*

In contrast, the best previous constructions, [MZ], had at least an exponential dependence on the degree $\ell$ in the seed-length. In particular, no non-trivial PRGs were known for fooling Lipschitz functions of degree $O(\log n)$ polynomials, and as a result for degree $O(\log n)$ PTFs, even for constant error rate.

As mentioned above, at a high-level, most of the previous PRGs for PTFs worked by first constructing a PRG for fooling smooth functions as above (with additional assumptions on the polynomials $P$) and then approximating the sign function suitably by smooth functions. Thus, our generator can be seen as a significant step towards obtaining nontrivial PRGs for PTFs of degree $O(\log n)$.

Besides the natural interest in constructing PRGs as in Theorem 1.2 in the context of fooling PTFs, an additional motivation for the question (which drew us to the problem in the first place) is the following application to hardness of approximating uniform sparsest cut.

## 1.1 Integrality Gaps for Uniform Sparsest Cut

*Uniform sparsest cut* is a fundamental problem in graph theory with a variety of applications in graph optimization and often appears as a basic step in several important approximation algorithms (for e.g., see the survey by Shmoys [Shm]).

Given an undirected graph $G = (V, E)$, the goal in the uniform sparsest cut problem is to find the subset of vertices $S \subseteq V$ that minimizes $|E(S, \overline{S})|/|S||\overline{S}|$, where $E(S, \overline{S})$ denotes the number of edges between $S$ and $\overline{S}$.

The uniform sparsest cut problem and its generalization–non-uniform sparsest cut[2], have been extensively studied in the context of approximation algorithms. The exact problem is known to be NP-hard [SM], but no NP-hardness results are known for approximating the optimum. On the other hand, constant-factor hardness of approximation results are known for non-uniform sparsest cut version assuming the unique games conjecture [CKK+, KV].

Designing approximation algorithms for sparsest cut has received a lot of attention in algorithm design, culminating in the seminal work of Arora, Rao and Vazirani [ARV] who gave a $O(\sqrt{\log n})$ factor approximation for uniform sparsest cut. This was later extended to non-uniform sparsest cut by Arora, Lee and Naor [ALN].

Arora, Rao and Vazirani in fact showed that an SDP relaxation for the problem first proposed by Goemans and Linial in the late 1990s, [Goe, Lin], has an integrality gap of $O(\sqrt{\log n})$ for uniform sparsest cut. Since then, understanding the performance of the Goemans-Linial SDP (so-called "basic SDP with triangle constraints") has been of much interest. A significant step in this direction was taken by the breakthrough work of Khot and Vishnoi [KV] who showed that the SDP relaxation for non-uniform sparsest cut has an integrality gap of at least $(\log \log n)^{1/6-o(1)}$, thus disproving a conjecture of Goemans and Linial. Following this, Devanur et al. [DKSV] showed that the Goemans-Linial SDP has a gap of $\Omega(\log \log n)$ even for uniform sparsest cut. Finally, Cheeger, Kleiner and Naor [CKN] showed an integrality gap of $(\log n)^{\Omega(1)}$ for non-uniform sparsest cut.

---

[1]Throughout, for a multi-set $S$, $x \sim S$ denotes the uniform distribution over $S$ and $\|P\| = \mathbb{E}_{x \sim \{1,-1\}^n}[P(x)^2]^{1/2}$.

[2]Here, one allows *demands* on the edges; we do not define the problem formally as we do not use it later on.

In this work we obtain a near-exponential improvement in the integrality gap of the Goemans-Linial SDP for uniform sparsest cut. (We refer to Section 3 for the formal definition of the SDP).

**Theorem 1.3.** *The Goemans-Linial SDP relaxation for uniform sparsest cut has an integrality gap of at least* $\exp\left(\Omega(\sqrt{\log \log n})\right)$.

We remark that the only better gap instance in this vein of Cheeger, Kleiner and Naor for non-uniform sparsest cut is known to *not* work for uniform sparsest cut. This difference between uniform sparsest cut and non-uniform sparsest cut is particularly interesting in the context of metric embedding, which we explain below.

## 1.2 Embedding Negative-type Metrics into $\ell_1$

There is a strong connection between embedding metrics into $\ell_1$ and the sparsest cut problem as was first evidenced in the work of Linial, London and Rabinovich [LLR]. By now it is well understood (for e.g., see [Rab], [CKN]) that the integrality gap of the Goemans-Linial SDP correspond exactly to the distortion of embedding negative-type metrics[3] into $\ell_1$. In particular, integrality gaps for the Goemans-Linial SDP for non-uniform sparsest cut correspond to *worst-case* distortion of such embeddings and integrality gaps for uniform sparsest cut correspond to *average-distortion* of such embeddings. Cheeger, Klein and Naor [CKN] show that the worst-case distortion can be as large as $(\log n)^{\Omega(1)}$. However, as mentioned in their work, their metric does in fact have a small average-distortion embedding into $\ell_1$. Our gap instance immediately gives the following corollary.

**Corollary 1.4.** *There exists a negative-type metric on n points that requires an average distortion of at least* $\exp\left(\Omega(\sqrt{\log \log n})\right)$ *to embed into* $\ell_1$.

## 1.3 Hierarchy Lower bounds for Sparsest Cut

Finally, we remark that our improved gap instance also translates to improvements in *hierarchy* lower bounds for uniform sparsest cut. SDP hierarchies are one of the most powerful techniques in algorithm design and knowing their limitations for specific problems often gives strong unconditional evidence for the hardness of the problem at hand (see [CT] for a recent survey). This is even more compelling for problems where we do not have NP-hardness results as is the case for uniform sparsest cut. One such important class of hierarchies is the Sherali-Adams hierarchy, which we augment here by starting with the *basic SDP relaxation*. Indeed, the Goemans-Linial SDP relaxation for sparsest cut is contained within a constant number of levels of this hierarchy.

---

[3]A metric space $(X, d)$ is said to be of negative type if it embeds isometrically into $\ell_2^2$: there exist $f : X \to \ell_2^k$ such that $d(x, y) = \|f(x) - f(y)\|_2^2$.

We defer a more precise description of the hierarchy to Section 3. Our lower bound for uniform sparsest cut in this hierarchy is as follows.

**Theorem 1.5.** *The integrality gap after R rounds of the Sherali-Adams hierarchy for uniform sparsest cut starting with the basic SDP relaxation is at least* $\exp\left(\Omega(\sqrt{\log \log n})\right)/R$.

The above result gives $\omega(1)$ integrality gap for $R = \exp\left(\Omega(\sqrt{\log \log n})\right)$ rounds. In contrast, the best previous results (even for the harder non-uniform sparsest cut problem) of Khot and Saket [KS1] and Raghavendra and Steurer [RS2] had non-trivial guarantees for at most $(\log \log n)^{\Omega(1)}$ rounds.

## 2 Proof Outline

### 2.1 PRGs for Lipschitz Functions of Polynomials

The basic generator we use is similar to that of Rabani and Shpilka [RS1] and Meka and Zuckerman [MZ]. However, our analysis is quite different and is arguably simpler. In particular, we do not need to appeal to the invariance principle for low-degree polynomials of Mossel, O'Donnel and Oleszkiewicz [MOO].

The generator we consider is the following. Fix $n, t, m = n/t$ and let $\mathcal{H} = \{h : [n] \to [t]\}$ be a family of hash functions. Let $\delta > 0$ be a parameter to be chosen later and let $G_h : \{0, 1\}^{r_1} \to \{1, -1\}^m$ be a PRG that fools halfspaces with error at most $\delta$. Let $G_{\mathcal{H}, G_h} : \{0, 1\}^r \to \{1, -1\}^n$ be the generator which samples a hash function $h \sim \mathcal{H}$, partitions the coordinates into buckets according to $h$ and uses an independent sample generated according to $G_h$ to set the coordinates in each bucket. We remark that the generator is similar to those in Rabani and Shpilka and Meka and Zuckerman with a different choice of $G_h$. Let $F : \{0, 1\}^s \to \{1, -1\}^m$ be a $2\ell$-wise independent family chosen independently of $G_{\mathcal{H}, G_h}$. We define a new generator $G_{\mathcal{H}, G_h, F}$ by taking the component-wise product of $G_{\mathcal{H}, G_h}$ and $F$.

We show that $G_{\mathcal{H}, G_h, F}$ fools Lipschitz functions of degree $\ell$ polynomials with error $\varepsilon$ when there are $t = O(\ell^2/\varepsilon^2)$ blocks and the inner halfspace generator $G_h$ fools halfspaces with error $\delta = O(\varepsilon^2/\ell)$. The analysis of the generator proceeds as follows.

Let $\psi$ be a Lipschitz function and $P$ a degree $\ell$ polynomial. Call a monomial in $P$ *bad* if the random hash $h \sim \mathcal{H}$ assigns more than one of the coordinates in the monomial to a single bucket. Delete all such bad monomials to get a polynomial $P_h$. We use pairwise independence of the hash family $\mathcal{H}$ to argue that any fixed monomial is bad with small probability. In particular, the total *weight* of deleted monomials in $P_h$ will be small in expectation. Finally, we use a simple tail bound to argue that as $\psi$ is Lipschitz, the error due to going from $\psi(P(\ ))$ to $\psi(P_h(\ ))$ will be small.

Finally, note that $P_h$ is linear within the coordinates in each bucket and we know that $G_h$ fools halfspaces with sufficiently small error. Combining this with a hybrid argument across the $t$ buckets shows that for any fixed hash function $h$, the generator $G_{\mathcal{H}, G_h, F}$ fools $\psi(P_h(\ ))$ with error at most $\delta \cdot t$. By setting the parameters appropriately and averaging over the choice of the hash functions $h \sim \mathcal{H}$ gives us the desired result. Theorem 1.2 then follows from using the PRG for halfspaces of Meka and Zuckerman [MZ] as $G_h$ in $G_{\mathcal{H}, G_h, F}$.

## 2.2 Integrality Gaps for Uniform Sparsest Cut

We next describe our results on uniform sparsest cut. The starting point for our improved gap instances is the *short code graph* (aka *Reed-Muller graph*) of Barak et al. [BGH+], who gave an exponentially smaller gadget (the 'short code') which can be used in place of *long code* in certain hardness reductions. At a high-level, our gap instance for uniform sparsest cut is obtained by replacing the long code with the short code in the construction of Devanur et al. [DKSV]. We analyze our gap instance using the framework of Raghavendra and Steurer [RS2].

The gap instance of Devanur et al. is obtained by looking at the hypercube (viewed as a long code) and *folding* the graph along cyclic shifts. That is, by collapsing sets of vertices which are cyclic shifts of one another to a single vertex. Our graph on the other hand is obtained by looking at the short code graph of Barak et al. whose vertices correspond to elements of the Reed-Muller code.

Unfortunately, the short code graph unlike the hypercube is not invariant under cyclic shifts. We get around this hurdle by observing that for the high-level intuition behind the folding operation of Devanur et al. to work, one only needs a group of automorphisms on the graph that are transitive on the *dictator* functions corresponding to the vertices. For Devanur et al., cyclic shifts satisfy this requirement. For the Reed-Muller graph of Barak et al. one such set of group actions is provided by *affine shifts*, where a $\mathbb{F}_2$ polynomial $P : \mathbb{F}_2^n \to \mathbb{F}_2$ is mapped to the polynomial $P_a(x) = P(x+a)$, for $a \in \mathbb{F}_2^n$. Thus, our final gap instance for uniform sparsest cut is obtained by folding (i.e., collapsing the vertices of) the short code graph of Barak et al. along the orbits of the above affine shift action.

The approach of replacing the long code with short code to obtain stronger integrality gap instances is already present in [BGH+] and in fact lead to significantly better integrality gap instances for unique games and various related constraint satisfaction problems. Unfortunately, making it work for the case of uniform sparsest cut (or even non-uniform sparsest cut) presents significant technical challenges and lead to no substantial quantitative improvements. In particular, a major bottleneck in the work of Barak et al. was the error parameter in a derandomized *majority is stablest* result over the Reed-Muller code shown by the authors. Using our improved PRG, we are able to show a much stronger (in fact exponentially better) majority is stablest result over the Reed-Muller code. This quantitative improvement provides the basis for the analysis of our integrality gap instances.

We remark that even after obtaining the better PRG as in Theorem 1.2 it is technically challenging to make sure all the pieces of [DKSV], [BGH+] and [RS2] fit together to give our final gap instance for uniform sparsest cut. In particular, we have to redo several technical claims from [BGH+] and [RS2] adapted to our specific context. We defer further details to the appropriate sections.

# 3 Preliminaries

## 3.1 Pseudorandomness

We use the following standard notions from pseudorandomness.

**Definition 3.1.** *A distribution $\mathcal{D}$ on $\{1,-1\}^n$ is k-wise independent if for every $I \subseteq [n]$ with $|I| \leq k$, and $x \sim \mathcal{D}$, the random variables $((x_i)_{i \in I})$ are independent and uniform over $\{1,-1\}$.*

**Definition 3.2.** *A family of hash functions $\mathcal{H} = \{h : [n] \to [t]\}$ is pairwise independent if for all $i \neq j \in [n]$ the random variables $h(i), h(j), h \sim \mathcal{H}$ are independent and uniform over $[t]$.*

There explicit distributions $\mathcal{D}$ and hash families $\mathcal{H}$ as above that can be sampled with $O(k \log n)$ [AS] and $O(\log n)$ [CW] random bits respectively.

We say a distribution $\mathcal{D}$ $\varepsilon$-fools a class of functions $\mathcal{F} = \{f : \{1, -1\}^n \to \{1, -1\}\}$ if for every $f \in \mathcal{F}$, $\mathbb{P}_{x \sim \mathcal{D}}[f(x) = 1] = \mathbb{P}_{x \sim \{1,-1\}^n}[f(x) = 1] \pm \varepsilon$. We shall use the following results of Diakonikolas et al. [DGJ$^+$] and Meka and Zuckerman [MZ] about fooling halfspaces (these are functions of the form $f(x) = \text{sign}(\langle w, x \rangle - \theta)$ for $w \in \mathbb{R}^n$ and $\theta \in \mathbb{R}$).

**Theorem 3.3** ([DGJ$^+$])**.** *There exists a constant $c$ such that for every $\varepsilon > 0$ and $k \geq c \log^2(1/\varepsilon)/\varepsilon^2$, $k$-wise independent distributions fool halfspaces with error at most $\varepsilon$.*

**Theorem 3.4** ([MZ])**.** *There exists an explicit PRG $G : \{0, 1\}^r \to \{1, -1\}^n$ that fools $\varepsilon$-fools halfspaces with a seed-length of $r = O(\log n + \log^2(1/\varepsilon))$.*

We also use the following simple large-deviation bound for variables with limited independence (see appendix for proof).

**Lemma 3.5.** *Let $X_1, \ldots, X_N \in \{1, -1\}$ be $k$-wise independent for $k$ even. Then, for all $t \geq 0$, $\mathbb{P}[|\sum_i X_i| \geq t \sqrt{N}] \leq k^{k/2}/t^k$.*

## 3.2 The Short Code

We next review the *short code graph* or *Reed-Muller graph* of Barak et al. [BGH$^+$]. Throughout, the Reed-Muller code will correspond to the Reed-Muller code over $\mathbb{F}_2$.

**Definition 3.6.** *The n-variate Reed-Muller code of degree $d$, denoted $\mathsf{RM}(n, d)$ is a length $2^n$ linear code with messages corresponding to n-variate polynomials $P$ from $\mathbb{F}_2^n \to \mathbb{F}_2$ of degree at most $d$. The encoding of a polynomial $P$ is $(P(x))_{x \in \mathbb{F}_2^n}$.*

We shall often, without explicitly stating so, view $\mathsf{RM}(n, d)$ as a subset of $\{1, -1\}^{2^n}$ via the mapping $a \in \{0, 1\} \leftrightarrow (-1)^a \in \{1, -1\}$. The meaning will be clear from the context.

We shall repeatedly use the following standard facts of Reed-Muller codes.

**Lemma 3.7.** *The dual of $\mathsf{RM}(n, d)$ is $\mathsf{RM}(n, n - d - 1)$ and the uniform distribution over $\mathsf{RM}(n, d)$ is $2^d$-wise independent.*

We next abstract the main properties of the short code graph from [BGH$^+$]. For fixed values of $n, d, N = 2^n$ and $\alpha \in \mathbb{F}_2^N$, let $\chi_\alpha : \mathbb{F}_2^N \to \{1, -1\}$ be the *character* defined by $\chi_\alpha(x) = (-1)^{\langle \alpha, x \rangle}$ and define
$$\deg(\chi_\alpha) = \min\{ wt(\alpha + y) : y \in \mathsf{RM}(n, n - d - 1) \}.$$

Barak et al. give a (weighted) Cayley graph whose vertex set is a Reed-Muller code and more importantly, whose spectrum closely approximates the spectrum of the *Boolean noisy cube*.

**Lemma 3.8** (See [BGH$^+$])**.** *For all $\varepsilon \in (0, 1/8)$, $n, d > 0$, $\rho = e^{-\varepsilon}$, $N = 2^n$, the following holds. There exists a weighted Cayley graph $G = G(n, d, \varepsilon)$ with vertex set $V = \mathsf{RM}(n, d)$ such that:*

- *The eigenvectors of $G$ are the characters $\{\chi_\alpha : \alpha \in \mathbb{F}_2^N/\mathsf{RM}(n, n-d-1)\}$[4]. Let $\lambda_\alpha$ denote the eigenvalue of character $\chi_\alpha$*

- *The graph $G$ is affine-shift invariant in the following sense. For any $b \in \mathbb{F}_2^n$, let $\pi_b : \mathsf{RM}(n,d) \to \mathsf{RM}(n,d)$ be defined by $\pi_b(P) = P'$ where for $P \in \mathsf{RM}(n,d)$, $P'$ is the polynomial with $P'(x) = P(x+b)$. Then, $\pi_b$ is an automorphism on $G$.*

- *For any $\alpha$, if $\deg(\chi_\alpha) = k$, $\lambda_\alpha \leq \max(\rho^{k/2}, \rho^{\mu_0 2^d})$ where $\mu_0$ is an absolute constant.*

- *For all $\delta < \delta_0$ for some constant $\delta_0$, if $\deg(\chi_\alpha) = k < \delta^2 2^{d+1}$, then $|\lambda_\alpha - \rho^k| \leq \delta$.*

- *For any vertex $u \in \mathsf{RM}(n,d)$ and $v$ a random neighbor of $u$ in $G$, $\mathbb{E}[\langle u, v \rangle] \geq (1-\varepsilon)N$ and for any two adjacent $u, v \in \mathsf{RM}(n,d)$, $\langle u, v \rangle > 3N/4$[5].*

## 3.3 Uniform Sparsest Cut

We next review some basics about the uniform sparsest cut problem along with the basic SDP relaxations. We will focus on the *balanced separator* problem; It is well known that integrality gap instances for balanced separator translate to similar gaps for uniform sparsest cut (see [DKSV] for instance).

Throughout this work we shall view graphs as given by a normalized adjacency matrix and will often view them as specified by (and specifying) a random walk on the set of vertices. Given a graph $G$ and a subset $S$ of vertices, let the conductance of $S$, $\phi_G(S)$, be the probability that a random edge out of a randomly chosen vertex of $S$ lands outside of $S$. Let $\rho_G$ denote the stationary distribution of $G$.

**Definition 3.9** (Balanced Separator Problem). *Given an undirected graph $G$ and a parameter $b \in (0, 1/2)$, find*
$$\phi(G, b) = \min\{\phi_G(S) : \rho_G(S) \in [b, 1-b]\}.$$

For our purposes it suffices to study the question where $b$ is any fixed constant, say $b = 1/3$.

We next describe the *standard* semi-definite relaxation of balanced separator problem with several rounds of the Sherali-Adams hierarchy. For a distribution $\mu$ on $\mathbb{R}^m$ and $A \subseteq [m]$, let $\mathsf{margin}_A(\mu)$ denote the marginal distribution of $\mu$ on the coordinates in $A$. Finally, for two distributions $\mu, \nu$, let $\|\mu - \nu\|_1$ denote the statistical distance between them. The SDP relaxation we consider is given in Figure 3.3. Intuitively, the relaxation can be seen as starting with the "standard" SDP for balanced separator and placing all possible *local integrality* constraints on any set of size at most $R$. We refer to [RS2] for more detailed motivation for the hierarchy. For our purposes, it mainly suffices to say that the Goemans-Linial SDP considered by Arora, Rao and Vazirani [ARV] lies within a constant number of levels of $\mathsf{SA}_R$ as described in Figure 3.3.

Given a collection of vectors $(v_i : i \in |V(G)|)$ and integral distributions $(\mu_S : S \subseteq [n], |S| \leq R)$ we call the pair $\mathsf{SA}_R$-*feasible* for balanced separator if together they satisfy the last four constraints of the SDP in Figure 3.3.

---

[4]This follows as $G$ is a Cayley graph and we have to quotient out the dual of $\mathsf{RM}(n,d)$ which is $\mathsf{RM}(n, n-d-1)$.

[5]Strictly speaking, the graph of Barak et al. has edges $\{u, v\}$ with $\langle u, v \rangle < 3N/4$, albeit with exponentially small weights. We enforce this condition as it can be done without any change to the theorem as stated and it avoids some annoying technicalities.

$\mathsf{SA}_R$-Hierarchy. Input: Graph $G$, $b \in (0, 1/2)$, $R$ - number of rounds. Variables of the SDP are in bold and $\rho_G$ denotes the stationary distribution on $G$.

$$\begin{aligned}
\text{minimize} \quad & \mathop{\mathbb{E}}_{(i,j)\sim E(G)} \frac{1}{4} \|\boldsymbol{v}_i - \boldsymbol{v}_j\|^2 \\
\text{subject to} \quad & \mathop{\mathbb{E}}_{(i,j)\sim \rho_G} \frac{1}{4} \|\boldsymbol{v}_i - \boldsymbol{v}_j\|^2 \geq 2b(1-b) \\
& \langle \boldsymbol{v}_i, \boldsymbol{v}_j \rangle = \mathop{\mathbb{E}}_{x \sim \boldsymbol{\mu}_S} [x_i x_j], && S \subseteq V(G), |S| \leq R,\ i, j \in S, \\
& \langle \boldsymbol{v}_i, \boldsymbol{v}_0 \rangle = \mathop{\mathbb{E}}_{x \sim \boldsymbol{\mu}_S} [x_i], && S \subseteq V(G), |S| \leq R,\ i \in S, \\
& \|\mathsf{margin}_{A \cap B}(\boldsymbol{\mu}_A) - \mathsf{margin}_{A \cap B}(\boldsymbol{\mu}_B)\|_1 = 0, && A, B \subseteq V(G), |A|, |B| \leq R, \\
& \boldsymbol{\mu}_S \text{ a distribution on } \{1, -1\}^S && S \subseteq V(G), |S| \leq R.
\end{aligned}$$

Figure 1: SDP relaxation of balanced separator in Sherali-Adams hierarchy

We shall use the results of Raghavendra and Steurer [RS2] and Khot and Saket [KS1] that show how to lift gap instances for the basic SDP to higher rounds of the Sherali-Adams hierarchy. We will mainly follow the framework of Raghavendra and Steurer. Below we state one of their results with a view towards our application to the balanced separator problem.

**Definition 3.10** ([RS2]). *A nice system of clouds is a collection $\mathcal{B}$ of subsets of $\mathbb{R}^d$ with the following properties:*

- *Every set $B \in \mathcal{B}$ consists of $N$ unit vectors. The sets $B \in \mathcal{B}$ are referred to as clouds.*

- *Near Orthogonality: For every $B \in \mathcal{B}$, and every unit vector $u \in \mathbb{R}^d$, $\sum_{v \in B} \langle u, v \rangle^2 \leq 3/2$.*

- *Matching Property: For every pair of clouds $(A, B) \in \mathcal{B}$, there exists a matching $M : A \to B$ such that for every $u \in A$, $M(u) = \mathrm{argmax}_{v \in B} |\langle u, v \rangle|$.*

- *Integrality: All vectors in $\mathcal{B}$ are elements of $\{\lambda, -\lambda\}^d$ for some fixed $\lambda$.*

The following result of Raghavendra and Steurer says that given a nice system of clouds one can get feasible solutions for several rounds of the Sherali-Adams hierarchy whose "geometry" (especially mutual correlations) is close to that of the clouds.

**Theorem 3.11** ([RS2]). *For $R, t > 0$, let $\delta = 10 \cdot R^2 \cdot e^{-t/16R}$. Then, for every nice system of clouds $\mathcal{B}$ over $\mathbb{R}^d$ of size $N$ each, there exists a $\mathsf{SA}_R$-feasible pair of vectors $(\boldsymbol{v}_B : B \in \mathcal{B})$ and distributions $(\boldsymbol{\mu}_S \text{ over } \{1, -1\}^S : S \subseteq \mathcal{B})$ such that the following holds. The vectors $(\boldsymbol{v}_B : B \in \mathcal{B})$ are of the form*

$$\boldsymbol{v}_B = \sqrt{1-\delta} \cdot \mathsf{normal}\left( \frac{1}{\sqrt{N}} \sum_{u \in B} u^{\otimes t} \right) + \sqrt{\delta} \cdot \boldsymbol{u}_B^\perp,$$

*where[6] $(\boldsymbol{u}_B^\perp : B \in \mathcal{B})$ is a set of unit vectors orthogonal to the vectors $\{u^{\otimes t} : u \in \cup B\}$.*

---

[6] For a vector $w \in \mathbb{R}^m$, $\mathsf{normal}(w)$ denotes the unit vector in the direction of $w$.

# 4 PRGs for Lipschitz Functions of Polynomials

We now give a PRG for Lipschitz functions of low-degree polynomials, proving Theorem 1.2. To avoid some minor technicalities, throughout we shall assume that the Lipschitz functions we consider are smooth in which case $\|\psi\|_{Lip} = \sup_x |\psi'(x)|$.

Recall the *hashing generator*, $G_{\mathcal{H},G_h,F}$, defined in the introduction, Section 2.1. We show that for suitable setting of $t, \delta$, $G_{\mathcal{H},G_h}$ fools Lipschitz functions of polynomials.

**Lemma 4.1.** *For every $\varepsilon > 0$, $t = 16\ell^2/\varepsilon^2$, $\delta \leq \varepsilon^4/64\ell^2$, $\mathcal{H}$ pairwise independent, the following holds for $G \equiv G_{\mathcal{H},G_h,F} : \{0,1\}^r \to \{0,1\}^n$. For every degree $\ell$ polynomial $P : \mathbb{R}^n \to \mathbb{R}$ with $\|P\| = 1$, and $\psi : \mathbb{R} \to \mathbb{R}$ a smooth function with bounded $\|\psi\|_{Lip}$,*

$$\left| \mathop{\mathbb{E}}_{x \in_u \{1,-1\}^n} [\psi(P(x))] - \mathop{\mathbb{E}}_{y \in_u \{0,1\}^r} [\psi(P(G(y)))] \right| < \|\psi\|_{Lip} \cdot \varepsilon.$$

In comparison, a similar result in [MZ] gets an error bound of $2^{O(\ell)} \cdot \varepsilon$. Also note that the above result requires no assumptions about $P$ and in particular $P$ need not be *regular* as required in [MZ].

*Proof.* Let $X \in_u \{1,-1\}^n$ and let $Y \equiv Y(h, Z^1, \ldots, Z^t, F)$ be the output of the generator, where $h \in_u \mathcal{H}$ and $Z^i$ denotes the samples generated from $G_h$ to set the variables in bucket $i$, and $F$ the output of our $2\ell$-independent family. For brevity, let $\bar{Z} = (Z^1, \ldots, Z^t)$.

Fix a hash function $h : [n] \to [t]$, and call a subset $I \subseteq [n]$ *h-bad* if $\max_{j \in [t]} |I \cap h^{-1}(j)| > 1$. Let

$$P(x) = \sum_{I : |I| \leq \ell} a_I \prod_{i \in I} x_i,$$

with $\sum_I a_I^2 = 1$. Let $P_h : \mathbb{R}^n \to \mathbb{R}$ be the degree $d$ polynomial obtained by deleting all $h$-bad monomials in $P$: $P_h = \sum_{I : I \text{ not } h\text{-bad}} a_I \prod_{i \in I} x_i$. We first show that the randomness in $\bar{Z}$ is enough to fool the polynomial $P_h$.

**Claim 4.2.** *For any fixed hash function $h$,*

$$\left| \mathbb{E}[\psi(P_h(X))] - \mathop{\mathbb{E}}_{\bar{Z},F}[\psi(P_h(Y))] \right| < 4\|\psi\|_{Lip} \cdot \sqrt{t\delta}.$$

*Proof.* We begin by showing that for any fixed values of $h$ and $F$, and for any $s \in \mathbb{R}$ that:

$$|\mathbb{P}(P_h(X) \leq s) - \mathbb{P}(P_h(Y) \leq s)| \leq t\delta.$$

We will prove this claim by a hybrid argument. The main intuition is that $P_h$ is affine in the variables of a single bucket and so should be fooled by the PRG for halfspaces $G_h$. Fix a hash function $h$ and let random variable $Y^i = (X^1, \ldots, X^i, Z^{i+1}, \ldots, Z^t)$ for $0 \leq i \leq t$, where $X^i \in_u \{1,-1\}^m$ and are independent of one another. Note that $Y^0 \equiv Y$ and $Y^t \equiv X \in_u \{1,-1\}^n$. Further, $Y^{i-1}, Y^i$ differ only in the $i$'th bucket which is $X^i$ in $Y^{i-1}$ and $Z^i$ in $Y^i$.

Fix $i \in [t]$ and let $Z$ denote the variables not in the $i$'th block. Without loss of generality suppose that $B^i = \{j : h(j) = i\} = [m]$. Then, as a function of the variables in $B^i$, $P_h$ is affine:

$$P_h(x_1, \ldots, x_m, Z, F) = x_1 \cdot P_1(Z, F) + x_2 P_2(Z, F) + \cdots + x_m P_m(Z, F) + P_0(Z, F) := \ell_{Z,F}(x),$$

where $P_1, \ldots, P_m, P_0$ are polynomials in $Z$ and $F$ of degree at most $\ell + 1$.

By assumption on the generator $G_h$, we have that:
$$\left|\mathbb{P}(\ell_{Z,F}(X^i) \leq s) - \mathbb{P}(\ell_{Z,F}(Z^i) \leq s)\right| \leq \delta.$$

Thus for each $i$,
$$\left|\mathbb{P}(P_h(Y^i) \leq s) - \mathbb{P}(P_h(Y^{i+1}) \leq s)\right| \leq \delta.$$

Iterating the above $t$ times yields
$$\left|\mathop{\mathbb{P}}_{X \sim_u \{1,-1\}^n}(P_h(X) \leq s) - \mathop{\mathbb{P}}_{Y \sim G_{\mathcal{H},G_h,F}}(P_h(Y) \leq s)\right| \leq t\delta.$$

Since $F$ is $2\ell$-wise independent, so is $G_{\mathcal{H},G_h,F}$. Thus, by Chebychev's inequality, we have that for any $t$
$$\mathop{\mathbb{P}}_{X \sim_u \{1,-1\}^n}(|P_h(X)| \geq s), \mathop{\mathbb{P}}_{Y \sim G_{\mathcal{H},G_h,F}}(|P_h(Y)| \geq s) \leq \frac{1}{s^2}.$$

Therefore, for a fixed $h$, by partial integration we have (we assume $\psi$ is bounded),
$$\left|\mathbb{E}\left[\psi(P_h(X))\right] - \mathbb{E}\left[\psi(P_h(Y))\right]\right| = \left|\int_{-\infty}^{\infty} \psi'(t) \cdot (\mathbb{P}[P_h(X) > s] - \mathbb{P}[P_h(Y) > s]) \, ds\right|$$
$$\leq \|\psi\|_{Lip} \cdot \int_{-\infty}^{\infty} |\mathbb{P}[P_h(X) > s] - \mathbb{P}[P_h(Y) > s]| \, ds$$
$$\leq \|\psi\|_{Lip} \int_{-\infty}^{\infty} \min\left(t\delta, s^{-2}\right) ds$$
$$\leq 2\|\psi\|_{Lip} \int_0^{1/\sqrt{t\delta}} t\delta \, dt + 2\|\psi\|_{Lip} \int_{1/\sqrt{t\delta}}^{\infty} s^{-2} ds$$
$$\leq 4\|\psi\|_{Lip} \cdot \sqrt{t\delta}.$$

$\square$

We now show that for a random $h \sim \mathcal{H}$, $P$ and $P_h$ are close.

**Claim 4.3.** $\left|\mathbb{E}_{Y,h}[\psi(P(Y))] - \mathbb{E}_{Y,h}[\psi(P_h(Y))]\right| < \|\psi\|_{Lip} \cdot (\ell/\sqrt{t})$.

*Proof.* Observe that for any $y \in \mathbb{R}^n$,
$$|\psi(P(y)) - \psi(P_h(y))| < \|\psi\|_{Lip} \cdot |P(y) - P_h(y)|.$$

Now, for any $I \subseteq [n]$, $\mathbb{P}_h[I \text{ is } h\text{-bad}] \leq \ell^2/t$. Therefore, for any fixed hash function $h$, as $Y$ is $2\ell$-wise independent, we get
$$\left|\mathbb{E}[\psi(P(Y))] - \mathbb{E}[\psi(P_h(Y))]\right| \leq \|\psi\|_{Lip} \cdot \mathop{\mathbb{E}}_h \mathop{\mathbb{E}}_Z [|P(Y) - P_h(Y)|]$$
$$\leq \|\psi\|_{Lip} \cdot \left(\mathop{\mathbb{E}}_h \mathop{\mathbb{E}}_Z \left[|P(Y) - P_h(Y)|^2\right]\right)^{1/2} = \|\psi\|_{Lip} \cdot \left(\mathop{\mathbb{E}}_h \left[\sum_{I: I \text{ is } h\text{-bad}} a_I^2\right]\right)^{1/2}$$
$$= \|\psi\|_{Lip} \cdot \left(\sum_{I: |I| \leq d} a_I^2 \cdot \mathop{\mathbb{P}}_h[I \text{ is } h\text{-bad}]\right)^{1/2} \leq \|\psi\|_{Lip} \cdot \frac{\ell}{\sqrt{t}}.$$

$\square$

Note that the above can also be applied to $X \in_u \{1, -1\}^n$. Therefore, combining the above two claims, we get

$$\left|\mathbb{E}_X[\psi(P(X))] - \mathbb{E}_Y[\psi(P(Y))]\right| \leq \left|\mathbb{E}_{X,h}[\psi(P(X))] - \mathbb{E}_{X,h}[\psi(P_h(X))]\right| +$$
$$\left|\mathbb{E}_X[\psi(P_h(X))] - \mathbb{E}_Y[\psi(P_h(Y))]\right| + \left|\mathbb{E}_{Y,h}[\psi(P_h(X))] - \mathbb{E}_{Y,h}[\psi(P(Y))]\right| \leq$$
$$2\|\psi\|_{Lip} \cdot (\ell/\sqrt{t}) + 4\|\psi\|_{Lip} \cdot \sqrt{t}\delta.$$

The lemma now follows by setting $t = 16\ell^2/\varepsilon^2$, $\delta = \varepsilon^4/64\ell^2$. □

Theorem 1.2 follows from the lemma and the PRG for halfspaces of Meka and Zuckerman [MZ].

*Proof of Theorem 1.2.* Using the PRG for halfspaces of Theorem 3.4 as $G_h$, the generator $G_{\mathcal{H}, G_h, F}$ as in Lemma 4.1 has a seed-length of $r = O(t \cdot r_h) = O((\log n + \log^2(\ell/\varepsilon))\ell^2/\varepsilon^2)$. □

## 5 Majority is Stablest over Reed-Muller Codes

As mentioned in the introduction, our integrality gap instance for uniform sparsest cut builds on the short code gadgets of Barak et al. [BGH+]. To do so, we first use our PRG for Lipschitz functions of polynomials to obtain a significant quantitative strengthening of the *majority is stablest over Reed-Muller codes* result of Barak et al. which in turn was based on the influential *majority is stablest* result of Mossel et al. [MOO]. We refer the reader to [MOO] for the motivation and history behind such results. The high-level structure of our majority is stablest result and its analysis are similar to those of Barak et al. who worked with the PRG for PTFs of Meka and Zuckerman [MZ]. However, the actual argument is somewhat delicate and we need to rework some technical arguments of Barak et al. We defer the details to the appendix.

To state our result we first define the notion of *influence* to functions defined over the Reed-Muller code. For a function $f : \mathsf{RM}(n, d) \to \mathbb{R}$ and $\alpha \in \mathbb{F}_2^N/\mathsf{RM}(n - d - 1)$ define the *Fourier coefficient* $\hat{f}(\alpha) = \mathbb{E}_{x \sim \mathsf{RM}(n,d)}[f(x) \cdot \chi_\alpha(x)]$.

**Definition 5.1.** *For a function $f : \mathsf{RM}(n, d) \to \mathbb{R}$ and $i \in [N]$, where $N = 2^n$ and $\ell > 0$, the $\ell$-degree influence of coordinate $i$ in $f$ is defined by*

$$\mathrm{Inf}_i^{\leq \ell}(f) = \sum_{\substack{\alpha \in \mathbb{F}_2^N/\mathsf{RM}(n,n-d-1), \\ \deg(\alpha) \leq \ell, \, \alpha_i = 1}} \hat{f}(\alpha)^2.$$

We can now state our majority is stablest over Reed-Muller codes result. For $\mu \in [0, 1]$, let $t(\mu) \in \mathbb{R}$ be such that $\mathbb{P}_{X \sim \mathcal{N}(0,1)}[X < t(\mu)] = \mu$. Then, for $\rho > 0$ define the *Gaussian stability curve* $\Gamma_\rho : \mathbb{R} \to \mathbb{R}$ by $\Gamma_\rho(\mu) = \mathbb{P}_{X,Y}[X \leq t(\mu), Y \leq t(\mu)]$, where $(X, Y) \in \mathbb{R}^2$ is a two-dimensional mean zero random Gaussian vector with covariance matrix $\begin{pmatrix} 1 & \rho \\ \rho & 1 \end{pmatrix}$.

**Theorem 5.2.** *There exist universal constants $c, C$ such that the following holds. Fix $n, d > 0$ and $\varepsilon > 0$. Let $G = G(n, d, \varepsilon)$ be the short code graph as in Lemma 3.8. Let $f\colon \mathsf{RM}(n, d) \to [0, 1]$ be a function on $\mathsf{RM}(n, d)$ with $\mathbb{E}_{x \sim \mathsf{RM}(n,d)}[f(x)] = \mu$ and $\max_{i \in [N]} \mathrm{Inf}_i^{\leq \log(1/\tau)}(f) \leq \tau$. Then, for $d > C \log \log(1/\tau)$,*

$$\mathbb{E}_{x \sim \mathsf{RM}(n,d)}[f(x) G f(x)] \leq \Gamma_\rho(\mu) + \frac{c \log \log(1/\tau)}{(1-\rho) \log(1/\tau)}, \tag{5.1}$$

*where $\rho = e^{-\varepsilon}$ and $\Gamma_\rho\colon \mathbb{R} \to \mathbb{R}$ is the noise stability curve of Gaussian space.*

Qualitatively this is similar to the statement of Barak et al. [BGH+]. However, quantitatively the above result is exponentially stronger in the requirement on the degree $d$ of the Reed-Muller code. Barak et al. require $d = \Omega(\log(1/\tau))$, whereas we only require $d = \Omega(\log \log(1/\tau))$. This improvement is critical for our improved integrality gap instances and could be of use in other applications of the short-code. We defer the proof to the appendix.

## 6  Integrality Gap Instances for Uniform Sparsest Cut

As mentioned in the introduction, roughly speaking, our gap instance is obtained by replacing the long code with the short code in the construction of Devanur et al. [DKSV]. We bound the integral value of the instance using Theorem 5.2 and bound the SDP value by applying the framework of Raghavendra and Steurer [RS2].

Fix a degree parameter $d$ and $\varepsilon > 0$ to be chosen later and $N = 2^n$. Let $G \equiv G(n, d, \varepsilon)$ denote the short code graph as in Lemma 3.8. Our candidate gap instance is obtained by *folding $G$* along orbits of the affine shift action. To this end, let $H$ be a subgroup of permutations of $[N]$ acting on $\mathbb{F}_2^N$ such that the following properties hold:

- $\mathsf{RM}(n, d)$ is closed under the action of $H$: For every $v \in \mathsf{RM}(n, d)$, and $\pi \in H$, $\pi(v) \in \mathsf{RM}(n, d)$ ($\pi(v)$ is the string obtained by permuting $v$ according to $\pi$).

- $H$ is transitive: For all $a, b \in [N]$, $\exists \pi \in H$ such that $\pi(a) = b$.

- $H$ acts as automorphisms on $G$: For all $u, v \in \mathsf{RM}(n, d)$, and $\pi \in H$, the weight of the edge $\{u, v\}$ in $G$ is the same as the weight of the edge $\{\pi(u), \pi(v)\}$ in $G$.

Given such a subgroup of permutations, define the *H-folded* graph $G_H$ in the natural way by collapsing the orbits $\{\pi(v) : \pi \in H\}$ into a single vertex for every $v \in \mathsf{RM}(n, d)$. Specifically, for any adjacent $u, v$ in $G$ with an edge-weight of $w_G(u, v)$ we add an edge of weight $w_G(u, v)$ between the orbits $\{\pi(u) : \pi \in H\}$ and $\{\pi(v) : \pi \in H\}$ in $G_H$.

For Reed-Muller codes, a natural group of permutations $H$ as above is given by affine shifts: for every $a \in \mathbb{F}_2^n$, consider the permutation $\pi$ on $[N]$ (recall that $N = 2^n$ and we identify $[N]$ with $\mathbb{F}_2^n$) given by the operation $x \to x + a$. It is easy to check that $H$ satisfies properties (1) and (2) above. Property (3) follows from Lemma 3.8. We will show that the graph $G_H$ has a large integrality gap for the balanced separator problem– Theorem 6.10.

In the following, we denote vertices of $G_H$ by bold letters $\boldsymbol{v}$, which we interpret as corresponding to orbits $\{\pi(v) : \pi \in H\}$ for vertices $v \in G$. Further, as $G$ is regular, to sample from the stationary distribution of $G_H$ it suffices to sample $v \sim \mathsf{RM}(n, d)$ and look at the corresponding orbit $\{\pi(v) : \pi \in G_H\}$. We state this below for later use.

**Fact 6.1.** *For $G_H$ as defined above, the stationary distribution on $G_H$ is the same as the distribution of orbits $\{\pi(v) : \pi \in G_H\}$, where $v \sim \mathsf{RM}(n, d)$.*

## 6.1 Bounding the Integral Value

We first bound the value of the integral solution of balanced separator on $G_H$. We do so by appealing to Theorem 5.2.

**Lemma 6.2.** *There exists a constant $C$ such that the following holds. For $d \geqslant C \log \log N$ and $\varepsilon = C((\log \log N)/\log N)^{2/3}$ and $t = \varepsilon 2^{d+1}$ the following holds. For $G \equiv G(n, d, \varepsilon)$ as in Lemma 3.8 and $H$ as defined above, every $1/3$-balanced cut in $G_H$ cuts at least $\Omega(\sqrt{\varepsilon})$ fraction of the edges.*

*Proof.* Let $\rho$ be the stationary distribution on $G_H$. For $b \in (0, 1/2)$, let $f' : V(G_H) \to \{0, 1\}$ define a $b$-balanced cut on $G_H$, i.e., $\rho(\{v : f'(v) = 1\}) \in [b, 1 - b]$. Lift the function $f'$ to all of $\mathsf{RM}(n, d)$ in the natural way as follows: define $f : \mathsf{RM}(n, d) \to \{0, 1\}$ by $f(v) = f'(v)$. From Fact 6.1, it follows that $f$ is $b$-balanced:

$$\mathbb{E}_{v \sim \mathsf{RM}(n,d)}[f(v)] = \mu \in [1/3, 2/3].$$

Further, observe that the fractional weight of edges cut by $f'$ in $G_H$ is exactly the fractional weight of edges cut by $f$ in $G$. We next show that $f$ has small influences in order to apply Theorem 5.2.

Let $D = 2^d$. Recall that for $\ell < D/2$ and $i \in [N]$, by Definition 5.1,

$$\mathsf{Inf}_i^{\leqslant \ell}(f) = \sum_{\alpha \in \mathbb{F}_2^N,\, |\alpha| \leqslant \ell,\, \alpha_i = 1} \hat{f}(\alpha)^2,$$

where $\hat{f}(\alpha) = \mathbb{E}_{x \sim \mathsf{RM}(n,d)}[f(x) \cdot \chi_\alpha(x)]$.

We will show that $\mathsf{Inf}_i^{\leqslant \ell}(f) = \mathsf{Inf}_j^{\leqslant \ell}(f)$ for all $i, j \in [N]$. Fix $i \neq j \in [N]$ and let $\pi \in H$ be such that $\pi(i) = j$ (such a permutation exists as $H$ is transitive). Then,

$$\begin{aligned}
\hat{f}(\pi(\alpha)) &= \mathbb{E}_{x \sim \mathsf{RM}(n,d)}[f(x) \cdot \chi_{\pi(\alpha)}(x)] \\
&= \mathbb{E}_{x \sim \mathsf{RM}(n,d)}[f(\pi(x)) \cdot \chi_{\pi(\alpha)}(\pi(x))] \\
&= \mathbb{E}_{x \sim \mathsf{RM}(n,d)}[f(x) \cdot \chi_\alpha(x)] \quad (f \text{ is constant on orbits under } H) \\
&= \hat{f}(\alpha).
\end{aligned}$$

Therefore,

$$\mathsf{Inf}_i^{\leqslant \ell}(f) = \sum_{\alpha \in \mathbb{F}_2^N,\, |\alpha| \leqslant \ell,\, \alpha_i = 1} \hat{f}(\alpha)^2 = \sum_{\alpha \in \mathbb{F}_2^N,\, |\alpha| \leqslant \ell,\, \alpha_i = 1} \hat{f}(\pi(\alpha))^2 = \sum_{\alpha \in \mathbb{F}_2^N,\, |\alpha| \leqslant \ell,\, \alpha_i = 1,\, \beta = \pi(\alpha)} \hat{f}(\beta)^2$$

$$= \sum_{\beta \in \mathbb{F}_2^N,\, |\beta| \leqslant \ell,\, \beta_j = 1} \hat{f}(\beta)^2 = \mathsf{Inf}_j^{\leqslant \ell}(f).$$

Further, note that for any function $f$,

$$\sum_{i \in [N]} \mathsf{Inf}_i^{\leqslant \ell}(f) \leqslant \sum_i \sum_{\alpha \in \mathbb{F}_2^N,\, |\alpha| \leqslant \ell,\, \alpha_i = 1} \hat{f}(\alpha)^2 \leqslant \ell \cdot \sum_\alpha \hat{f}(\alpha)^2 \leqslant \ell.$$

Therefore, $\text{Inf}_i^{\leq \ell}(f) \leq \ell/N$.

Let $\ell = \log N$, $\tau = (\log N)/N$. Then, from the above equation $\text{Inf}_i^{\leq \log(1/\tau)} \leq \tau$. Thus, by Theorem 5.2 applied to $f$, for $d \geq C \log \log(1/\tau)$ and $\varepsilon$ small enough so that $1 - \varepsilon < e^{-\varepsilon} = \rho < 1 - \varepsilon/2$,

$$\mathbb{E}_{x \sim \text{RM}(n,d)}[f(x)Gf(x)] \leq \Gamma_\rho(\mu) + \frac{c \log \log(1/\tau)}{(1-\rho)\log(1/\tau)} \leq 1 - \Omega(\sqrt{\varepsilon}) + \frac{O(\log \log N)}{\varepsilon \log N} \leq 1 - \Omega(\sqrt{\varepsilon}),$$

where the last-but-one inequality follows from known facts about $\Gamma_\rho$– see [MOO], [BGH$^+$], and the last inequality from setting the value of $\varepsilon$ as in the lemma.

The above bound immediately translates to a bound on the fraction of edges crossing the cut defined by $\{x : f(x) = 1\}$. In particular, $f$ cuts at least a $\Omega(\sqrt{\varepsilon})$ fraction of the edges in $G$. The theorem now follows as the conductance of the cut defined by $f'$ in $G_H$ is the same as the conductance of the cut defined by $f$ in $G$. □

## 6.2 Bounding the SDP Value

We next bound the value of the natural SDP relaxation for balanced separator as given in Figure 3.3 on the graph $G_H$. Most of this section is devoted to proving the following.

**Lemma 6.3** (Main SDP value). *There exists a constant $C$ such that the following holds. For $d \geq C \log \log N$ and $\varepsilon = C((\log \log N)/\log N)^{1/3}$, $R \leq (\log N)^{O(1)}$ the following holds. For $G, H$ as in Lemma 6.2, the R'th round of $\text{SA}_R$-hierarchy relaxation defined in Figure 3.3 for the balanced separator problem has objective value at most $O(R\varepsilon \log(1/\varepsilon))$.*

We prove the lemma by exhibiting a $\text{SA}_R$-feasible solution with low objective value. We do so by first constructing a nice system of clouds as in Theorem 3.11 and then applying the theorem to get a $\text{SA}_R$-feasible set of vectors and distributions. We then use the properties of $G_H$ and the system of clouds we construct to bound the objective value.

For $v \in \text{RM}(n, d) \subseteq \{1, -1\}^N$, let $B(v) \subseteq \mathbb{R}^{N^3}$, be the (possibly multi-)set of vectors

$$B(v) = \left\{ \left(\pi(v)/\sqrt{N}\right)^{\otimes 3} : \pi \in H \right\}. \tag{6.1}$$

Let $\mathcal{B} = (B(v) : v \in \text{RM}(n, d))$ denote all the clouds. We first show that by throwing away few of the clouds $B(v)$, we get a nice system of clouds in the sense of Theorem 3.11.

To this end, for $v \in \text{RM}(n, d)$, call the cloud $B(v)$ *nearly-orthogonal* if the following holds:

$$\max_{\pi \in H} |\langle v, \pi(v) \rangle| \leq N^{2/3}/2. \tag{6.2}$$

The following shows that a nearly-orthogonal cloud satisfies the near orthogonality property of Theorem 3.11.

**Claim 6.4.** *For any nearly orthogonal cloud $B(v) \in \mathcal{B}$, $\max_{w \in \mathbb{R}^{N^3}, \|w\|=1} \sum_{u \in B(v)} \langle w, u \rangle^2 \leq 3/2$.*

*Proof.* We can bound the max in the lemma in terms of the largest eigenvalue of the Gram matrix of the vectors $B(v)$, $M(v) \in \mathbb{R}^{N \times N}$ defined by $M(v)_{ab} = \langle \pi_a(v)^{\otimes 3}, \pi_b(v)^{\otimes 3} \rangle/N^3$ for $a, b \in \mathbb{F}_2^N$. Now, by (6.2), $M(v)_{ab} \leq 1/8N$ for $a \neq b$. Thus, the off-diagonal entries of $M(v)$ are at most $1/8N$ and the diagonal entries are 1. Thus, by Gershgorin theorem, the largest eigenvalue of $M(v)$ is at most $1 + (N-1)/8N \leq 9/8$. The claim now follows. □

We next argue that most of the clods $B(v) \in \mathcal{B}$ are nearly-orthogonal. This follows from a straightforward application of Lemma 3.5.

**Claim 6.5.** *For random $v \sim \mathsf{RM}(n, d)$, the cloud $B(v)$ defined by (6.1) is nearly-orthogonal with probability at least $1 - N^{-\Omega(2^d)}$.*

*Proof.* Fix any $a \in \mathbb{F}_2^n$. Then, for $v = (P(x))_{x \in \mathbb{F}_2^n}$,

$$\langle v, \pi_a(v) \rangle = \sum_{x \in \mathbb{F}_2^n} (-1)^{P(x)+P(x+a)}.$$

By Lemma 3.7, for $d \geqslant 2$, $a \neq 0$, and $P$ a random degree $d$ polynomial, the random variables $((-1)^{P(x)+P(x+a)})_{x \in \mathbb{F}_2^n}$ are $k$-wise independent for $k = 2^{d-1}$. Therefore, by Lemma 3.5, for $t = N^{1/6}/2$,

$$\mathop{\mathbb{P}}_{v \sim \mathsf{RM}(n,d)} \left[ |\langle v, \pi_a(v) \rangle| > N^{2/3}/2 \right] \leqslant k^{k/2}/t^k. \tag{6.3}$$

Thus, by a union bound, for $v \sim \mathsf{RM}(n, d)$ $B(v)$ is nearly-orthogonal with probability at least $1 - N \frac{k^{k/2}}{t^k} = 1 - N^{-\Omega(2^d)}$. □

We next show that the clouds $B(v)$ satisfy the matching property.

**Claim 6.6.** *For $u, v \in \mathsf{RM}(n, d)$, $B(u), B(v)$ satisfy the matching property.*

*Proof.* Let $\pi, \sigma \in H$ be such that $|\langle \pi(u), \sigma(v) \rangle| = \max_{u' \in B(u), v' \in B(v)} |\langle u', v' \rangle|$. Then, for any $\rho \in H$, $\langle \rho(\pi(u)), \rho(\sigma(v)) \rangle = \langle \pi(u), \sigma(v) \rangle$. Therefore, $B(u), B(v)$ satisfy the matching property, by considering the matching $\{(\rho(\pi(u)), \rho(\sigma(v))) : \rho \in H\}$ between $B(u), B(v)$. □

It is clear that the clouds $B(u)$ satisfy the integrality property. Let $\mathcal{B}' = \{B(v) : v \in \mathsf{RM}(n, d), v \text{ is nearly-orthogonal as in } (6.2)\}$. Then, from the above arguments, $\mathcal{B}'$ satisfies the hypothesis of Theorem 3.11. Let $(v_B : B \in \mathcal{B}')$ and $(\mu_S \text{ over } \{1, -1\}^S : S \subseteq \mathcal{B}')$ be the $\mathsf{SA}_R$-feasible pair of vectors and distributions as guaranteed by Theorem 3.11 for an odd integer value $t$ to be chosen later. Then, for $\delta = R^2 \exp(-\Omega(t/R))$ as in the theorem:

$$v_B = \sqrt{1 - \delta} \cdot \mathsf{normal}\left( \frac{1}{\sqrt{N}} \sum_{u \in B} u^{\otimes t} \right) + \sqrt{\delta} \cdot u_B^\perp, \tag{6.4}$$

where $(u_B^\perp : B \in \mathcal{B})$ is a set of unit vectors orthogonal to the vectors $\{u^{\otimes t} : u \in \cup B\}$. We thus have vectors and distributions for all nearly-orthogonal clouds. We extend these to all clouds of $\mathcal{B}$ arbitrarily: fix a good cloud $B_0$ and for every cloud $B \in \mathcal{B} \setminus \mathcal{B}'$, let $v_B = v_{B_0}$. As will be clear later, this will not effect any of our arguments quantitatively as most clouds are nearly-orthogonal by Claim 6.5.

The above arguments give us a vector for every cloud $B \in \mathcal{B}$. As clouds $B \in \mathcal{B}$ naturally correspond to orbits of the action of $H$ on $\mathsf{RM}(n, d)$, we get a vector for every vertex of the graph $G_H$. Similarly, the distributions $(\mu_S \text{ over } \{1, -1\}^S : S \subseteq \mathcal{B})$ naturally extend to distributions on subsets of vertices of $G_H$. We show that this collection of vectors and distributions give us a good feasible solution for the SDP in Figure 3.3 on $G_H$. In the following we shall freely translate between vertices of the graph $G_H$ and the clouds $B \in \mathcal{B}$ - the meaning will be clear from context.

Theorem 3.11 already shows that the above candidate solution satisfies the last four constraints of Figure 3.3. We next show that $(v_B)$ satisfies the *balance condition* (first constraint in Figure 3.3). To do so we shall use the following technical claim which helps us relate the correlation between vectors $v_{B(u)}, v_{B(v)}$ as defined by (6.1) to the correlation between $u, v$. We defer the proof to the appendix.

**Claim 6.7.** *Let $B(u), B(v)$ be nearly-orthogonal clouds. Then,*

$$\langle v_{B(u)}, v_{B(v)} \rangle = \frac{1}{N^{3t}} \sum_{\sigma \in H} \langle u, \sigma(v) \rangle^{3t} \pm N^{-\Omega(t)} \pm O(\delta). \tag{6.5}$$

*Further, if in addition $\langle u, v \rangle \geq 3N/4$, then*

$$\langle v_{B(u)}, v_{B(v)} \rangle \geq (\langle u, v \rangle/N)^{3t} - 2^{-\Omega(t)} - O(\delta). \tag{6.6}$$

**Claim 6.8** (Balance property). *Let $\rho$ be the stationary distribution on $G_H$. Then, for $2^d \geq 3t$, and vectors $(v_B)_{B \in \mathcal{B}}$ as defined by (6.4), $\mathbb{E}_{B, B' \sim \rho} \|v_B - v_{B'}\|^2 / 4 \geq 1/2 - O(\delta) - N^{-\Omega(2^d)} - N^{-\Omega(t)}$.*

*Proof.* We start by observing that by Fact 6.1, the distribution of $(B, B')$ for $B, B' \sim \rho$ is the same as that of $(B(u), B(v))$ for $u, v \sim \mathsf{RM}(n, d)$. We then focus on near-orthogonal clouds $B(u), B(v)$ as most clouds satisfy this property by Claim 6.5. For nearly-orthogonal clouds $B(u), B(v)$, we use Claim 6.7 to bound the correlation between $v_{B(u)}, v_{B(v)}$ by looking at correlations between vectors $u, v$. Finally, we use the fact that $u, v \sim \mathsf{RM}(n, d)$ are $2^d$-wise independence to show that they are almost orthogonal to one another.

Let $u, v \sim \mathsf{RM}(n, d)$ and let $\mathcal{E}$ denote the event that $u, v$ are nearly-orthogonal. By Claim 6.5, $\mathbb{P}[\mathcal{E}] \geq 1 - N^{-\Omega(2^d)}$. Further, by Lemma 3.7, $u, v$ are $2^d$-wise independent as strings over $\{1, -1\}$. Thus, for $3t < 2^{d-1}$ and $t$ odd, for any fixed $\sigma \in H$,

$$\begin{aligned} 0 = \mathop{\mathbb{E}}_{u', v' \sim \{1, -1\}^N} \left[ \langle u', \sigma(v') \rangle^{3t} \right] &= \mathbb{E} \left[ \langle u, \sigma(v) \rangle^{3t} \right] \\ &= \mathbb{P}[\mathcal{E}] \cdot \mathbb{E} \left[ \langle u, \sigma(v) \rangle^{3t} \mid \mathcal{E} \right] + \mathbb{P}[\neg \mathcal{E}] \mathbb{E} \left[ \langle u, \sigma(v) \rangle^{3t} \mid \neg \mathcal{E} \right] \\ &= \mathbb{P}[\mathcal{E}] \cdot \mathbb{E} \left[ \langle u, \sigma(v) \rangle^{3t} \mid \mathcal{E} \right] \pm N^{-\Omega(2^d)} \cdot N^{3t}. \end{aligned}$$

Therefore, $\mathbb{E}\left[\langle u, \sigma(v) \rangle^{3t} \mid \mathcal{E}\right] \geq -N^{-\Omega(2^d)} \cdot N^{3t}$. Combining this with (6.5), we get

$$\mathbb{E}\left[\langle v_{B(u)}, v_{B(v)} \rangle \mid \mathcal{E}\right] \geq -N^{-\Omega(2^d)} - N^{-\Omega(t)} - O(\delta).$$

Finally, as all of the vectors $v_{B(u)}$ are unit vectors, we get,

$$\mathbb{E}\left[\langle v_{B(u)}, v_{B(v)} \rangle\right] \geq \mathbb{P}[\mathcal{E}] \cdot \mathbb{E}\left[\langle v_{B(u)}, v_{B(v)} \rangle \mid \mathcal{E}\right] - \mathbb{P}[\neg \mathcal{E}] \geq -N^{-\Omega(2^d)} - N^{-\Omega(t)} - O(\delta).$$

Therefore,

$$\mathop{\mathbb{E}}_{B, B' \sim \rho} \|v_B - v_{B'}\|^2 = 2 - 2 \mathop{\mathbb{E}}_{B, B' \sim \rho} \langle v_B, v_{B'} \rangle = 2 - 2 \mathop{\mathbb{E}}_{u, v \sim \mathsf{RM}(n, d)} \langle v_{B(u)}, v_{B(v)} \rangle \geq 2 - O(\delta) - N^{-\Omega(2^d)} - N^{-\Omega(t)}.$$

The claim now follows. □

We next bound the objective value for the set of vectors $(v_B : B \in \mathcal{B})$. Recall that there is a natural association between vertices of the graph $G_H$ and the clouds of $\mathcal{B}$. In the following, for a cloud $B$, let $B' \sim_{G_H} B$ denote the cloud corresponding to a random neighbor in $G_H$ of the vertex associated with cloud $B$.

**Claim 6.9** (Objective value). *Let $\rho$ denote the stationary distribution on $G_H$. Then, for $\varepsilon \leq 1/4$ and vectors $(v_B)_{B \in \mathcal{B}}$ as defined by (6.4), $\mathbb{E}_{B \sim \rho,\, B' \sim_{G_H} B} \|v_B - v_{B'}\|^2 \leq O(t\varepsilon) + O(\delta) + 2^{-\Omega(t)} + N^{-\Omega(2^d)}$.*

*Proof.* We first simplify our task by moving from the folded graph $G_H$ to the original graph $G$. From Fact 6.1, the distribution of $(B, B')$ for $B \sim \rho$ and $B' \sim_{G_H} B$ is the same as that of $(B(u), B(v))$, where $u \sim \mathsf{RM}(n,d)$ and $v \sim_G u$ (i.e., $v$ is a random neighbor of $u$ in $G$). Thus, to prove the claim, it suffices to bound $\mathbb{E}_{u \sim \mathsf{RM}(n,d),\, v \sim_G u} \|v_{B(u)} - v_{B(v)}\|^2$.

The high-level argument is now similar to that of the proof of Claim 6.8. We focus on near-orthogonal clouds $B(u)$ as most clouds are nearly-orthogonal. We then argue that for nearly-orthogonal clouds $B(u), B(v)$ with $v \sim_G u$, $\langle B(u), B(v) \rangle$ is large as $\langle u, v \rangle$ is large. The last fact follows from the properties of the short code graph $G$.

Let $u \sim \mathsf{RM}(n,d)$ and $v \sim_G u$. By Lemma 3.8, $\langle u, v \rangle \geq 3N/4$. Let $\mathcal{E}$ be the event that $u, v$ are nearly-orthogonal. Then, by (6.6),

$$\mathbb{E}\left[\langle v_{B(u)}, v_{B(v)} \rangle \mid \mathcal{E}\right] \geq \mathbb{E}\left[(\langle u,v \rangle/N)^{3t} \mid \mathcal{E}\right] - 2^{-\Omega(t)} - O(\delta)$$
$$\geq \mathbb{E}\left[(\langle u,v \rangle/N)^{3t}\right] - 2\,\mathbb{P}[\neg\mathcal{E}] - 2^{-\Omega(t)} - O(\delta)$$
$$\geq (1-\varepsilon)^{3t} - N^{-\Omega(2^d)} - 2^{-\Omega(t)} - O(\delta)$$
$$\geq 1 - O(t\varepsilon) - N^{-\Omega(2^d)} - 2^{-\Omega(t)} - O(\delta),$$

where the last-but-one inequality follows from Lemma 3.8 and the power-mean inequality. Therefore, by Claim 6.5,

$$\mathbb{E}\left[\langle v_{B(u)}, v_{B(v)} \rangle\right] \geq \mathbb{P}[\mathcal{E}]\,\mathbb{E}\left[\langle v_{B(u)}, v_{B(v)} \rangle \mid \mathcal{E}\right] - \mathbb{P}[\neg\mathcal{E}] \geq 1 - O(t\varepsilon) - O(\delta) - 2^{-\Omega(t)} - N^{-\Omega(2^d)}.$$

Hence,
$$\mathbb{E} \|v_{B(u)} - v_{B(v)}\|^2 = 2 - 2\,\mathbb{E}\langle v_{B(u)}, v_{B(v)} \rangle = O(t\varepsilon) + O(\delta) + 2^{-\Omega(t)} + N^{-\Omega(2^d)}.$$

The claim now follows. □

We are now ready to prove the main claim of this section.

*Proof of Lemma 6.3.* The lemma follows from setting the parameters appropriately in Claims 6.8, 6.9. Let us use the same notations as in Claims 6.8, 6.9. Recall that in Theorem 3.11 $\delta = R^2 \exp(-\Omega(t/R))$. Therefore, by choosing $t = O(R \log(1/\varepsilon))$ sufficiently large, by Claim 6.9, we get

$$\mathbb{E}_{B \sim \rho,\, B' \sim_{G_H} B} \|v_B - v_{B'}\|^2 = O(R\varepsilon \log(1/\varepsilon)).$$

Similarly, for $C$ a sufficiently large constant, $2^d \geq 3t$. Thus, by Claim 6.8, we get

$$\mathbb{E}_{B, B' \sim \rho} \|v_B - v_{B'}\|^2 / 4 \geq 2b(1-b).$$

Therefore, the vectors $(v_B : B \in \mathcal{B})$ form a feasible solution for the SDP in Figure 3.3. The lemma follows. □

## 6.3 Gap Instance

We are now ready to state our main integrality gap result for balanced separator problem. This follows immediately from Lemmas 6.2, 6.3.

**Theorem 6.10.** *For all n, there exists a graph G on M-vertices such that for all R, the integrality gap for the balanced-separator problem on G after R rounds of $\mathsf{SA}_R$-hierarchy is at least*

$$\exp\left(\Omega(\sqrt{\log \log M})\right)/R.$$

*Proof.* Consider the graph $G_H$ as in Lemmas 6.2, 6.3. Then, as $d = C \log \log N$, the number of vertices $M$ in $G_H$ is

$$M \leq |\mathsf{RM}(n,d)| \leq \exp(n^d) = \exp\left((\log N)^d\right) = \exp\left(\exp\left((\log \log N)^2\right)\right).$$

Thus, $\log N = \Omega\left(\exp(\sqrt{\log \log M})\right)$. We can thus assume $R < \log N$, as else the claim becomes trivial.

Further, by Lemmas 6.2, 6.3, for $\varepsilon = \Omega(((\log \log N)/\log N)^{2/3})$, the integrality gap after $R$ rounds of the SDP in Figure 3.3 on $G_H$ is at least

$$\Omega\left(\frac{1}{R\varepsilon^{1/2}\log(1/\varepsilon)}\right) = \Omega\left(\frac{(\log N)^{1/3}}{R(\log \log N)^2}\right) = \Omega\left(\frac{(\log N)^{1/4}}{R}\right) = \Omega\left(\frac{\exp\left(\sqrt{\log \log M}/4\right)}{R}\right).$$

The theorem follows. □

The main integrality gap results stated in the introduction follow from the above theorem and the well-known relation between balanced separator and uniform sparsest cut (see [DKSV] for instance).

*Proof of Theorem 1.3.* Follows from fact that the Goemans-Linial SDP can be realized within a constant number of rounds of $\mathsf{SA}_R$-hierarchy and setting $R = O(1)$ in the above theorem. □

*Proof of Corollary 1.4.* Follows from the above theorem and the known equivalences between the integrality gaps for the Goemans-Linial SDP and distortion of embedding negative-type metrics into $\ell_1$. See [Rab], [CKN] for instance. □

**Acknowledgements.** We thank Prasad Raghavendra and David Steurer for valuable discussions.

## A   Majority is Stablest over Reed-Muller Codes

We now prove Theorem 5.2. To do so, we first use our PRG for Lipschitz functions to show an *invariance principle* for low-degree polynomials over the Reed-Muller code along the lines of the invariance principle for low-degree polynomials of Mossel et al. [MOO]. Specifically, we show a statement similar to Lemma 4.1 for the function $\zeta : \mathbb{R} \to \mathbb{R}$ defined by $\zeta(x) = (\min(0, x, 1 - x))^2$.

Let $G_{\mathcal{H},G_h}, G_{\mathcal{H},G_h,F}$ be as in the statement of Lemma 4.1. As shown by Barak et al. [BGH+], we use the fact that Reed-Muller codes of degree $D$ can be realized as instantiations of $G_{\mathcal{H},G_h}$ for $D \geq \log(kt)$ where $G_h$ generates a $k$-wise independent distribution on $\{1, -1\}^m$. Further, if $G_{\mathcal{H},G_h}$ were already a $2\ell$-wise independent family obtained by sampling uniformly from a vector subspace over $\mathbb{F}_2$, as is the case for Reed-Muller codes, we may take $F = G_{\mathcal{H},G_h}$ and thus have $G_{\mathcal{H},G_h,F} = G_{\mathcal{H},G_h}$.

Further, by [DGJ+], $k$-wise independent distributions fool halfspaces with error at most $O(\sqrt{(\log k)/k})$. Combining these observations with Lemma 4.1, we get the following invariance principle for the function $\zeta$.

**Theorem A.1.** *There exists a constant $c_3$ such that for every $\ell, \varepsilon > 0$, and $d \geq c_3 \log(\ell/\varepsilon)$ the following holds. For every degree $\ell$ polynomial $P : \mathbb{R}^n \to \mathbb{R}$ with $\|P\| = 1$, $n = 2^m$,*

$$\left| \mathbb{E}_{X \in_u \{1,-1\}^n} [\zeta(P(X))] - \mathbb{E}_{Y \in_u \mathsf{RM}(m,d)} [\zeta(P(Y))] \right| < \varepsilon.$$

*Proof.* Let $\psi(x) = \zeta(x) - x^2$. It should be noted that $|\psi|_{Lip} = 2$. Since the Reed-Muller code is $2\ell$-independent, we have that $\mathbb{E}_{X \in_u \{1,-1\}^n} [P(X)^2] = \mathbb{E}_{Y \in_u \mathsf{RM}(m,d)} [P(Y)^2]$. It thus suffices to show that

$$\left| \mathbb{E}_{X \in_u \{1,-1\}^n} [\psi(P(X))] - \mathbb{E}_{Y \in_u \mathsf{RM}(m,d)} [\psi(P(Y))] \right| < \varepsilon.$$

Let $k = \tilde{O}(\ell^2/\varepsilon^4)$ and $t = O(\ell^2/\varepsilon^2)$ be as in Lemma 4.1 so that $G_{\mathcal{H},G_h,F}$ fools $\psi(P(\ ))$ with error at most $|\psi|_{Lip} \cdot \varepsilon/2 = \varepsilon$ when $G_h$ generates a $k$-wise independent distribution.

Now, for $d \geq \log(kt) = O(\log(\ell/\varepsilon))$, $Y \in_u \mathsf{RM}(m,d)$ can be seen as an instance of $G_{\mathcal{H},G_h,F}$. This completes our proof. □

A similar result is shown by Barak et al. [BGH+], however the error guarantee they get has an exponential dependence on the degree $\ell$ of the polynomial $P$. This improvement is crucial for our applications.

The proof of Theorem 5.2 is similar to the one of Barak et al. [BGH+]. Unfortunately, we cannot use their proof as is, but need to rework their somewhat technical argument for technical reasons. The high-level idea is to use the invariance principle from the above result to *derandoimze* the result of Mossel et al. [MOO] for the *Boolean noisy cube*. For $\rho > 0$, let $T_\rho$ denote the Boolean noisy graph with vertices corresponding to $\{1,-1\}^N$ and two vertices $x, y \in \{1,-1\}^N$ have an edge of weight $\rho^{d_H(x,y)}(1-\rho)^{n-d_H(x,y)}$, where $d_H(\ ,\ )$ denotes the Hamming distance. We use the following lemma that follows easily from [MOO]. Below, we use *expectation inner products* for functions, i.e., for real-valued functions $f, g$ on a universe $U$, $f, g : U \to \mathbb{R}$, $\langle f, g \rangle = \mathbb{E}_{X \sim U}[f(x)g(x)]$.

**Lemma A.2.** *Let $f: \{1,-1\}^N \to \mathbb{R}$ be such that $\mathbb{E} f = \mu$, $\mathbb{E} f^2 \leq 1$ and $\mathbb{E} \zeta \circ f \leq \eta$. Suppose $\mathrm{Inf}_i^{\leq \log(1/\tau)} f \leq \tau$ for all $i \in [N]$. Then,*

$$\langle f, T_\rho f \rangle \leq \Gamma_\rho(\mu) + O(\eta) + O\left(\frac{1}{1-\rho}\right) \cdot \frac{\log \log(1/\tau)}{\log(1/\tau)}.$$

*where $T_\rho$ is the Boolean noise graph with second largest eigenvalue $\rho$ and $\Gamma_\rho$ is the Gaussian noise stability curve.*

*Proof of Theorem 5.2.* Let $\varepsilon = 1/\log^2(1/\tau)$, $\delta = 1/\log^2(1/\tau)$, $\ell = \log(1/\tau)$. Let $d = C \log \log(1/\tau)$ for $C$ sufficiently large to be chosen later so that (for $c_3$ as in Theorem A.1) $d \geq c_3 \log(\ell/\varepsilon)$, $\ell < \delta^2 2^{d+1}$.

For $\alpha \in \mathbb{F}_2^N/C$, let $\lambda_\alpha$ be the eigenvalues of $G$. Then, by Lemma 3.8,

$$|\lambda_\alpha - \rho^k| < \delta, \text{ for } k \leq \ell, \quad |\lambda_\alpha| < \rho^{\ell/2}, \text{ for } k > \ell. \tag{A.1}$$

Let $\gamma < 1/8$ be a parameter to be chosen later. Let $g = G^\gamma f$ and $G' = G^{1-2\gamma}$. Then, the graph $G'$ has the same eigenfunctions as $G$ - $\chi_\alpha$ for $\alpha \in \mathbb{F}_2^N/C$ with eigenvalues $\lambda'_\alpha = \lambda_\alpha^{1-2\gamma}$. From the above equation, it is easy to check that, for $\rho' = \rho^{1-2\gamma}$,

$$|\lambda'_\alpha - (\rho')^k| < \sqrt{\delta}, \text{ for } k \leq \ell, \quad |\lambda'_\alpha| < (\rho')^{\ell/2}, \text{ for } k > \ell. \tag{A.2}$$

Now, decompose $g = g^{\leq \ell} + g^{>\ell}$ into a *low-degree* part $g^{\leq \ell} = \sum_{\alpha \in \mathbb{F}_2^n, \text{wt}(\alpha) \leq \ell} \hat{g}(\alpha)\chi_\alpha$ and a *high-degree* part $g^{>\ell} = \sum_{\alpha \in \mathbb{F}_2^n/C, \Delta(\alpha, C) > \ell} \hat{g}(\alpha)\chi_\alpha$. Then,

$$\langle f, Gf \rangle = \langle g, G'g \rangle = \langle g^{\leq \ell}, G'g^{\leq \ell} \rangle + \langle g^{>\ell}, G'g^{>\ell} \rangle \leq \langle g^{\leq \ell}, G'g^{\leq \ell} \rangle + \mu \cdot \max_{\alpha \in \mathbb{F}_2^N/C, \Delta(\alpha, C) > \ell} \lambda'_\alpha.$$

Hence, using Equation (A.2) (and the crude bound $\mu \leq 1$),

$$\langle f, Gf \rangle \leq \sum_{\alpha \in \mathbb{F}_2^N, \text{wt}(\alpha) \leq \ell} (\rho')^{\text{wt}(\alpha)} \hat{g}(\alpha)^2 + (\rho')^\ell + \sqrt{\delta}. \tag{A.3}$$

In the remainder of this section we shall view $\mathsf{RM}(n, d)$ as a subset of $\{1, -1\}^N$. Then, as $g$ is $[0, 1]$-valued on $\mathsf{RM}(n, d)$ and $\zeta$ measures distance to bounded random variables, by Equation (A.1),

$$\mathbb{E}_{z \sim S}\left[\zeta\left(g^{\leq \ell}(z)\right)\right] \leq \mathbb{E}_{z \sim S}\left[\left(g(z) - g^{\leq \ell}(z)\right)^2\right] = \mathbb{E}_{z \sim S}\left[\left(g^{>\ell}(z)\right)^2\right] = \mathbb{E}_{z \sim S}\left[\left(G^\gamma f^{>\ell}(z)\right)^2\right] \leq \max_{\alpha: |\alpha|>\ell}(\lambda_\alpha^\gamma)^2 \leq \rho^{\gamma\ell}.$$

Hence, by Theorem A.1, (recall that $\ell = \log(1/\tau), \varepsilon = 1/\log^2(1/\tau)$)

$$\mathbb{E}_{x \sim \{1,-1\}^N}\left[\zeta\left(g^{\leq \ell}(x)\right)\right] \leq \mathbb{E}_{z \sim S}\left[\zeta\left(g^{\leq \ell}(z)\right)\right] + \varepsilon \leq \underbrace{\rho^{\gamma\ell} + \varepsilon}_{\eta :=}.$$

Now, as $\mathsf{RM}(n, d)$ is $\ell$-wise independent ($\ell < 2^{d+1}$),

$$\mathbb{E}_{x \sim \{1,-1\}^N}\left[g^{\leq \ell}(x)\right] = \mathbb{E}_{z \sim S}\left[g^{\leq \ell}(z)\right] = \mathbb{E}_{z \sim S}[g(z)] \pm \mathbb{E}_{z \sim S}\left[\left(g^{>\ell}(z)\right)^2\right]^{1/2} \leq \mu + \sqrt{\eta}.$$

Observe that $g^{\leq \ell}$ is a multilinear polynomial of degree at most $\ell$ and as the $\ell$-degree influences of $g$ are at most $\tau$, $g^{\leq \ell}$ is $\tau$-regular. Therefore, by Lemma A.2,

$$\langle g^{\leq \ell}, T_{\rho'} g^{\leq \ell} \rangle = \sum_{\alpha: \text{wt}(\alpha) \leq \ell} (\rho')^{\text{wt}(\alpha)} \hat{g}(\alpha)^2 \leq \Gamma_{\rho'}(\mu + \sqrt{\eta}) + O(\eta) + \frac{O(\log \log(1/\tau))}{(1-\rho)\log(1/\tau)}. \tag{A.4}$$

Since $\Gamma_{\rho'}(\mu + \sqrt{\eta}) \leq \Gamma_{\rho'}(\mu) + 2\sqrt{\eta}$ and $\Gamma_\rho(\mu) \leq \Gamma_{\rho'}(\mu) + |\rho - \rho'|/(1-\rho)$ (cf. Lemma B.3, Corollary B.5 in [MOO]), it follows from (A.3), (A.4) that

$$\langle f, Gf \rangle = \langle g, G'g \rangle \leq \Gamma_\rho(\mu) + O\left(\frac{|\rho - \rho'|}{1-\rho}\right) + O(\sqrt{\eta})\frac{O(\log\log(1/\tau))}{(1-\rho)\log(1/\tau)} + \rho^{(1-2\gamma)\ell} + \delta^{1/2}$$

$$= \Gamma_\rho(\mu) + \frac{O(\log\log(1/\tau))}{(1-\rho)\log(1/\tau)} + O\left(\frac{\gamma\log(1/\rho)}{1-\rho} + \rho^{\gamma\ell/2} + \varepsilon^{1/2} + \delta^{1/2}\right).$$

(Here we used the estimate $|\rho - \rho'| = |\rho - \rho^{1-2\gamma}| = O(\gamma \log(1/\rho))$.) Thus, for $\gamma = C \log\log(1/\tau)/\log(1/\tau)$ for sufficiently large $C$, the above expression simplifies to

$$\langle f, Gf \rangle \leq \Gamma_\rho(\mu) + \frac{O(\log\log(1/\tau))}{(1-\rho)\log(1/\tau)}.$$

This completes the proof of Theorem 5.2. □

# B Missing Proofs

*Proof of Lemma 3.5.* By Khintchine inequalities and the $k$-wise independence of $X_i$'s we get $\mathbb{E}[|\sum_i X_i|^k] \leqslant k^{k/2} N^{k/2}$. The claim now follows by Markov's inequality. $\square$

We now prove Claim 6.7 from Section 6. We use the following auxiliary claim.

**Claim B.1.** *For any nearly-orthogonal cloud $B(u)$, $\|\sum_{u' \in B(u)} (u')^{\otimes t}\|/\sqrt{N} \geqslant 1 - N^{-\Omega(t)}$.*

*Proof.* For any nearly-orthogonal cloud $B(u)$, and $u_1 \neq u_2 \in B(u)$, by (6.2), $|\langle u_1, u_2 \rangle| < (N^{-1/3})^3 = 1/N$. Therefore,

$$\left\|\sum_{u' \in B(u)} (u')^{\otimes t}\right\|^2 \geqslant \sum_{u' \in B(u)} \left\|(u')^{\otimes t}\right\|^2 - \sum_{u_1 \neq u_2 \in B(u)} |\langle u_1, u_2 \rangle|^t \geqslant N - N^2/N^t.$$

The claim now follows. $\square$

*Proof of Claim 6.7.* Fix two nearly-orthogonal clouds $B(u), B(v) \in \mathcal{B}$. By (6.1),

$$\left\langle \sum_{u' \in B(u)} (u')^{\otimes t}, \sum_{v' \in B(v)} (v')^{\otimes t} \right\rangle = \left\langle \sum_{\pi \in H} (\pi(u))^{\otimes 3t}/N^{3t/2}, \sum_{\sigma \in H} (\sigma(v))^{\otimes 3t}/N^{3t/2} \right\rangle$$

$$= \frac{N}{N^{3t}} \left\langle u^{\otimes 3t}, \sum_{\sigma \in H} (\sigma(v))^{\otimes 3t} \right\rangle = \frac{N}{N^{3t}} \sum_{\sigma \in H} \langle u, \sigma(v) \rangle^{3t}.$$

Now, by Claim B.1 and the above equations we get,

$$\left\langle \text{normal}\left(\sum_{u' \in B(u)} (u')^{\otimes t}\right), \text{normal}\left(\sum_{v' \in B(v)} (v')^{\otimes t}\right) \right\rangle = \left(1 \pm N^{-\Omega(t)}\right) \left\langle \sum_{u' \in B(u)} (u')^{\otimes t}, \sum_{v' \in B(v)} (v')^{\otimes t} \right\rangle$$

$$= \left(1 \pm N^{-\Omega(t)}\right) \frac{N}{N^{3t}} \sum_{\sigma \in H} \langle u, \sigma(v) \rangle^{3t}. \quad \text{(B.1)}$$

Finally, note that in (6.4) the vectors $u_B^\perp$ are orthogonal to all of the cloud vectors. Then, by (6.4) and (B.1)

$$\langle v_{B(u)}, v_{B(v)} \rangle \geqslant (1-\delta)\left(1 - N^{-\Omega(t)}\right) \frac{1}{N^{3t}} \sum_{\sigma \in H} \langle u, \sigma(v) \rangle^{3t} - \delta.$$

Equation (6.5) now follows.

Equation (6.6) follows from a similar claim in [RS2]. We omit the proof here. $\square$